\documentclass[12pt]{iopart}
\usepackage[dvips]{graphicx}

\begin{document}

\title{Double null formulation of the general Vaidya metric}

\author{Cecilia Chirenti$^1$ and Alberto Saa$^2$}

\address{$^1$Centro de Matem\'atica, Computa\c c\~ao e Cogni\c c\~ao, UFABC, 09210-170 Santo Andr\'e, SP, Brazil}

\address{$^2$Departamento de Matem\'atica Aplicada, UNICAMP,  13083-859 Campinas, SP, Brazil}

\eads{\mailto{asaa@ime.unicamp.br},\mailto{cecilia.chirenti@ufabc.edu.br}}

\date{\today}

\def \beq {\begin{equation}}
\def \eeq {\end{equation}}

\begin{abstract}
We present here the field equations describing a non-stationary spherically symmetric $n$-dimensional
charged black hole with varying mass $m(v)$ and/or electric charge
$q(v)$, described by a generic charged Vaidya metric with cosmological
constant $\Lambda$ in double null coordinates. This formulation of the metric has been shown to be particularly useful for perturbative studies and it was used in some recent works. Here we also discuss some issues related to the apparent and event horizons of the black hole.
\end{abstract}


\submitto{\CQG}

\maketitle

\section{Introduction}

In 1951, P. C. Vaidya presented for the first time a metric to describe the spacetime outside a spherically symmetric star, taking into account the radiation flux emitted by the star \cite{Vaidya}. In his work, the mass of the star is no longer considered to be constant and the metric is not static. The situation is described as a spherical mass surrounded by a finite and nonstatic envelope of radiation with radial symmetry. 

This metric has been the usual starting point for
the study of the quasinormal modes of time dependent black holes 
\cite{Xue:2003vs,Shao:2004ws,Abdalla:2006vb,Abdalla:2007hg,He:2009jd}. 
The charged version of the
metric is also an usual starting point for the study of many  aspects
of charged black hole physics
\cite{charge1,charge2,Ori91,Fayos,Parikh:1998ux,Hong,Hwang}.

The Vaidya metric has also been used to describe
 spherically symmetric collapse and the formation of
naked singularities \cite{Joshi,Lake}. 
It was also applied to the study of Hawking
radiation and the process of black hole evaporation
\cite{Hiscock,Kuroda,Biernacki,Parentani}, in the stochastic gravity
program \cite{Bei-Lok}, and in recent numerical relativity
investigations \cite{Nielsen:2010wq}. 

In the study of quasinormal modes, it is useful to write the Vaidya metric in double null coordinates \cite{Abdalla:2006vb,Abdalla:2007hg,Cecilia}. However, there is no general coordinate transformation from the usual radiation coordinates to double null coordinates \cite{WL}. So our purpose with this paper is to present the most general Vaidya metric, $n$-dimensional, with cosmological constant $\Lambda$ and with time dependent mass and electrical charge in these coordinates.

The numerical results are obtained with a generalization of the semi-analytical method used in \cite{GS,Saa:2007ej}, that allows us to construct the structure of the spacetime from the behavior of the outgoing null geodesics.

The structure of the paper is as follows. In section \ref{sec2} we present the derivation of the most general Vaidya metric in double-null coordinates. In section \ref{sec3} we present a discussion on the structure of the spacetime and the numerical results obtained with the semi-analytical method. And finally in section \ref{sec4} we present our final discussion of the results.

\section{The general Vaidya metric in double-null coordinates}
\label{sec2}

The $n$-dimensional Vaidya metric was first discussed in \cite{IV}. It
can be easily cast  in $n$-dimensional radiation coordinates
$(v,r,\theta_1,\dots,\theta_{n-2})$ as done, for instance, in
\cite{GD}. The $n$-dimensional charged Vaidya metric in radiation
coordinates, obtained originally in \cite{charge2}, reads 
\begin{equation}
\label{Vaidya}
\fl
\rmd s^2 = -\left(1-\frac{2m(v)}{(n-3)r^{n-3}}
+ \frac{2q^2}{(n-2)(n-3)r^{2(n-3)}}
\right)\rmd v^2 
+ 2c\rmd r\rmd v+ r^2\rmd\Omega^2_{n-2},
\end{equation}
 where $n>3$,  $c=\pm 1$, and $\rmd\Omega^2_{n-2}$ stands for the metric
 of the unit $(n-2)$-dimensional sphere, assumed here to be spanned by
 the angular coordinates $(\theta_1,  \theta_{2},\dots ,\theta_{n-2})$, 
\beq
\rmd\Omega^2_{n-2} = \sum_{i=1}^{n-2}\left(\prod_{j=1}^{i-1} \sin^2\theta_j \right) \rmd\theta_{i}^2.
\eeq
For the case of an ingoing radial flow, $c=1$ and $m(v)$ is a monotonically
increasing mass function in the advanced time $v$, while $c=-1$
corresponds to an outgoing radial flow, with $m(v)$ being in this case 
a monotonically decreasing mass function in the retarded time $v$. The
constant $q$ corresponds to the total electric charge. In principle,
one can also consider time dependent charges $q$ as done, for instance,
in \cite{Ori91}. This situation will of course require the presence of
charged null fluids and currents, whose realistic nature  we do not
address here. 

It has been known since a long time that the radiation
coordinates are defective at the horizon \cite{LSM}, implying that 
the Vaidya metric (\ref{Vaidya}), with or without electric charge, is
not geodesically complete in any dimension. The radiation coordinates are not enough to cover the entire Vaidya spacetime. (The radiation coordinates are defective at horizons where $v^2 \to \infty$). As can be seen in \cite{LSM}, for a 4-dimensional Vaidya metric with $dm/du < 0$ and without electric charge (in the context of ref. \cite{LSM}, a radiating star), the hypersurface $r = 2m(\infty)$ at $v = \infty$ in the Vaidya metric is analogous to the Schwarzschild hypersurface $r = 2m$ at Schwarzschild's time coordinate $T = +\infty$ in the Kruskal metric. (See \cite{Fayos} and
\cite{Booth:2010eu} for further discussions about possible analytical  
extensions and properties of the horizon of the Vaidya metric in
the radiation coordinates). 

The cross  term $drdv$
 introduces  extra terms in the hyperbolic equations governing
the evolution of physical fields on spacetimes with the metric
 (\ref{Vaidya}). Typically, the double-null coordinates are   far more
convenient for QNM analysis. This was the main motivation of the
series of works based on Waugh and Lake's approach \cite{WL}, where 
the problem of casting the 4-dimensional Vaidya metric in double-null
coordinates was originally addressed. As all previous attempts to
construct a general transformation from radiation to double-null
coordinates had failed, Waugh and Lake   considered the problem of
solving Einstein's equation with spherical symmetry directly in
double-null coordinates. The resulting equations, however, are not
analytically solvable in general. Waugh and Lake's
work was revisited in \cite{GS}, where a semi-analytical approach 
allowing for general mass functions was proposed. 

More recently 
this semi-analytical approach was extended to the
case of an $n$-dimensional Vaidya metric with cosmological constant
$\Lambda$ \cite{Saa:2007ej}. This  approach consists in a qualitative
study of the null-geodesics, allowing the description of light-cones
and revealing many features of the underlying causal structure.
It can also be used for more quantitative analyses; indeed, it
has already enhanced considerably the accuracy of the quasinormal
modes analysis of varying mass black holes
\cite{Abdalla:2006vb,Abdalla:2007hg}, and it can also be applied to
the study of gravitational collapse\cite{GS}. 

In this section, we extend the approach proposed in \cite{Saa:2007ej}
 and derive the double-null formulation for the most general Vaidya 
 metric:  $n$-dimensional, in the presence of a cosmological constant,
 and with  varying electric charge and mass.  Only the main results are
 presented. The reader can get more details on the employed
 semi-analytical approach in \cite{Saa:2007ej} and the references
 cited therein. We recall that the $n$-dimensional spherically
 symmetric line element in double-null coordinates
 $(u,v,\theta_1,\dots,\theta_{n-2})$ is given by
\beq
\label{uv}
\rmd s^2 = -2f(u,v)\rmd u\,\rmd v + r^2(u,v)\rmd\Omega^2_{n-2},
\eeq
where $f(u,v)$ and $r(u,v)$ are non vanishing smooth functions. The
energy-momentum tensor of a unidirectional radial null-fluid in the
eikonal approximation in the presence of an electromagnetic field
$F_{ab}$ is given by 
\beq
\label{T}
T_{ab} = \frac{1}{8\pi}hk_a k_b
+\frac{1}{4\pi}\left(F_{ac}F^{\phantom{b}c}_{b} -\frac{1}{4}g_{ab}
F_{cd}F^{cd}\right)
,
\eeq
where $k_a$ is a radial null vector and $h(u,v)$ is a smooth function
characterizing the null-fluid radial flow. In the double null coordinates $(u,v,\theta_1,\ldots,\theta_{n-2})$, we can choose either $k^a = (1,0,0,\ldots,0)$ (flow along the $u$-direction) or $k^a = (0,1,0,\ldots,0)$ (flow along the $v$-direction). Since the $u$ and $v$ directions are unspecified, it is not in fact necessary to consider flows along both directions. We will consider here,
without loss of generality, the case of a flow along the $v$-direction, as done in \cite{WL}.

Maxwell equations are given by
\numparts
\begin{eqnarray}
\label{Max}
\frac{1}{\sqrt{-g}}\partial_a \sqrt{-g}  F^{ab} = 4\pi J^b, \\
F_{ab,c} + F_{ca,b} + F_{bc,a} = 0\,,
\label{Maxwell2}
\end{eqnarray}
\endnumparts
where, for the metric (\ref{uv}),
\beq
\sqrt{-g} = fr^{n-2}\prod_{j=1}^{n-3}\left( \sin\theta_j\right)^{n-j-2}.
\eeq
All geometrical quantities relevant to this work are listed in the
appendix of  \cite{Saa:2007ej}. The equations (\ref{Max}) have the
following static spherically symmetric solution 
\beq
\label{staticsol}
F^{uv} = - F^{vu}  = \frac{q}{fr^{n-2}},
\eeq
with all other components of the electromagnetic tensor vanishing,
where $q$ is a constant which one identifies as  the $n$-dimensional
electric monopole charge. This case, of course, corresponds to
$J^b=0$. In order to allow for a time dependent charge $q(v)$, one
needs to assume the presence of a current
\beq
J^u = \frac{1}{4\pi}\frac{\dot{q}(v)}{fr^{n-2}},
\eeq
(with all other components vanishing) which is obtained from the continuity equation $J^a_{\phantom{a};a} = 0$, as done, for instance, in
\cite{Ori91}. Such a current must naturally appear, as we will see, in
the $h(u,v)$ function characterizing the radial flow in the
energy momentum tensor (\ref{T}). We note here that our solution (\ref{staticsol}) for $F^{ab}$ with $q(v)$ also consistently satisfies the sourceless Maxwell equations (\ref{Maxwell2}).

Einstein equations with  cosmological constant $\Lambda$
\beq
\label{eee}
R_{ab} - \frac{1}{2}g_{ab}R = -\Lambda g_{ab} + 8\pi T_{ab}\,,
\eeq
imply that, for the energy-momentum tensor (\ref{T}),
\beq
\label{contr}
R = \frac{2n}{n-2}\Lambda
-2\frac{n-4}{n-2}\frac{q^2}{r^{2(n-2)}}
,
\eeq
where (\ref{staticsol}) was used. The ${}_{uu}$ and ${}_{vv}$
components of Einstein equations for this case read simply 
\begin{eqnarray}
\label{eq1}
 \frac{f_{,u}}{f} - \frac{r_{,uu}}{r_{,u}} = 0, \\
\label{eq2}
 \frac{f_{,v}}{f} - \frac{r_{,vv}}{r_{,v}}  =  \frac{h}{n-2} \frac{r}{r_{,v}},
\end{eqnarray}
where $_{,u}$ and $_{,v}$ denote, respectively,
differentiation with respect to $u$ and $v$ as usual. The ${}_{uv}$ and
${}_{\theta_j\theta_j}$ components are, respectively, 
\begin{eqnarray}
\label{eq3}
 \frac{f_{,u}f_{,v}}{f^2} - \frac{f_{,uv}}{f} - (n-2)\frac{r_{,uv}}{r} =  -\frac{2\Lambda}{n-2} f   + 2 \frac{n-3}{n-2}f\frac{q^2}{r^{2(n-2)}},  \\
\label{eq4}
 \frac{2}{f} \left(rr_{,uv} + (n-3)r_{,u}r_{,v} \right) + (n-3)  = \frac{2\Lambda}{n-2} r^2 +
\frac{2 }{n-2}\frac{q^2}{r^{2(n-3)}}.
\end{eqnarray}
For $n\ne 3$, differentiating Eq. (\ref{eq4}) with respect to $u$ and
then inserting Eq. (\ref{eq1}) leads to
\beq
\label{A}
\frac{r^{n-2}r_{,uv}}{f} - \frac{2\Lambda}{(n-2)(n-1)}r^{n-1}
-\frac{2}{n-2}\frac{q^2}{r^{n-3}}
= -m,
\eeq
(after integration with respect to $u$) where $m(v)$ is an  arbitrary integration function that we already
known from \cite{Saa:2007ej} that can be interpreted as the mass of
the solution. The $n=3$ case  must be considered separately, in an
analogous way as done for $q=0$ in  \cite{Saa:2007ej}, and the most important results are presented in the Appendix. Now, differentiating Eq. (\ref{eq4}) with
respect to $v$ and using (\ref{eq2}) and (\ref{A}) gives
\beq
\label{hh}
h = -\left(\frac{n-2}{n-3}\right)  \frac{f}{r^{n-2}r_{,u}}
\left(m_{,v} - \frac{1}{n-2}\frac{(q^2)_{,v}}{r^{n-3}} \right)
.
\eeq
Eq. (\ref{eq1}) is ready to be integrated
\beq
\label{B}
f = 2Br_{,u}\,,
\eeq
where $B(v)$ is another arbitrary (but nonvanishing) integration
function. From (\ref{hh}) and (\ref{B}), one has
\beq
\label{h}
h = -2\left(\frac{n-2}{n-3}\right)  \frac{B}{r^{n-2}}
\left(m_{,v} - \frac{1}{n-2}\frac{(q^2)_{,v}}{r^{n-3}} \right).
\eeq
Finally, by using (\ref{eq4}) and (\ref{B}), Eq. (\ref{A}) can be
written as 
\beq
\label{r2}
\fl
r_{,v} = -B \left(1 - \frac{2m(v)}{(n-3)r^{n-3}} - \frac{2\Lambda}{(n-2)(n-1)}r^2 + \frac{2}{(n-2)(n-3)}\frac{q^2(v)}{r^{2(n-3)}}
\right).
\eeq
Note that (\ref{eq3}) and (\ref{B}) reproduce
(\ref{A}). Einstein equations are, therefore, equivalent to the
equations (\ref{B}), (\ref{h}), and (\ref{r2}), generalizing the
previous results of \cite{WL}, \cite{GS} and \cite{Saa:2007ej}. 

As already mentioned, the physical interpretation of  the arbitrary
integration functions $m(v)$ and $B(v)$ are the same of the $q=0$
case. Transforming from the double-null coordinates back to the
radiation coordinates by the coordinate  change $(u,v)\rightarrow
(r(u,v),v)$, the metric (\ref{uv}) will read
\beq
\label{uv1}
\rmd s^2 = 4Br_{,v} \rmd v^2 - 4B\rmd r\rmd v + r^2\rmd\Omega^2_{n-2},
\eeq
where (\ref{B}) was explicitly used. Comparing (\ref{Vaidya}) and
(\ref{uv1}) and taking (\ref{r2}) into account, it is clear that with
the choice $B = \pm 1/2,$ the function $m(v)$ indeed represents the
mass of the $n$-dimensional charged solution. The coordinate transformation leading to (\ref{uv1}) also ensures that the Vaidya metric in radiation coordinates (\ref{Vaidya}) and the double null metric (\ref{uv}) constructed in this paper are (locally) isometric. It is important to stress this fact, given the absence of a Birkhoff theorem for non vacuum spacetimes.

As in the $q=0$ case, the weak energy condition applied for (\ref{T}) requires, from (\ref{h}), that 

\begin{equation}
B\left(m_{,v} - \frac{1}{n-2}\frac{(q^2)_{,v}}{r^{n-3}} \right)< 0\,.
\label{energycond}
\end{equation}

If there are both mass and charge variations, then $m_{,v}$ and $q_{,v}$ cannot be chosen arbitrarily (see \cite{Ori91} for a discussion) and must chosen satisfy the energy condition (\ref{energycond}).

Taking (as mentioned above) $B = \pm \frac{1}{2}$, if we have only mass (or charge) varying with time, the energy condition requires that $m(v)$ (or $q(v)$) be a monotonic function and fixes $B$ in the following way:

\begin{eqnarray}
\textrm{if}\ m_{,v} > 0 \ (\textrm{or} \ q_{,v} < 0) \ \textrm{then} \ B = -\frac{1}{2}\,, \nonumber \\
\textrm{if}\ m_{,v} < 0 \ (\textrm{or} \ q_{,v} > 0) \ \textrm{then} \ B = +\frac{1}{2}\,, \nonumber
\end{eqnarray}

where we consider, without loss of generality, $q(v) > 0$.

\section{The spacetime structure}
\label{sec3}

The problem of constructing a double-null formulation for the general
Vaidya metric may be stated as follows: given the mass function
$m(v)$, the electric charge function $q(v)$, the cosmological constant
$\Lambda$, and the constant $B$, one needs to solve Eq. (\ref{r2}),
obtaining the function $r(u,v)$. Then, $f(u,v)$ and $h(u,v)$ are
calculated from (\ref{B}) and (\ref{h}). The arbitrary function  of
$u$ appearing in the integration of (\ref{r2}) must be chosen
properly \cite{WL}, so that $f(u,v)$ given in (\ref{B}) is a non-vanishing function.
 Unfortunately, as stressed previously by Waugh and
Lake\cite{WL}, such a procedure is not analytically solvable in
general. In \cite{Saa:2007ej},   a semi-analytical procedure is
introduced to attack the problem of solving Eqs. (\ref{B})-(\ref{r2})
for the $q=0$ case, generalizing in this way the results of \cite{GS}
obtained for $n=4$ and $\Lambda=0$. 
\begin{figure}[!htb]
\vspace{-0.5cm}
\begin{center}
  \includegraphics[angle=270,width=1.0\linewidth]{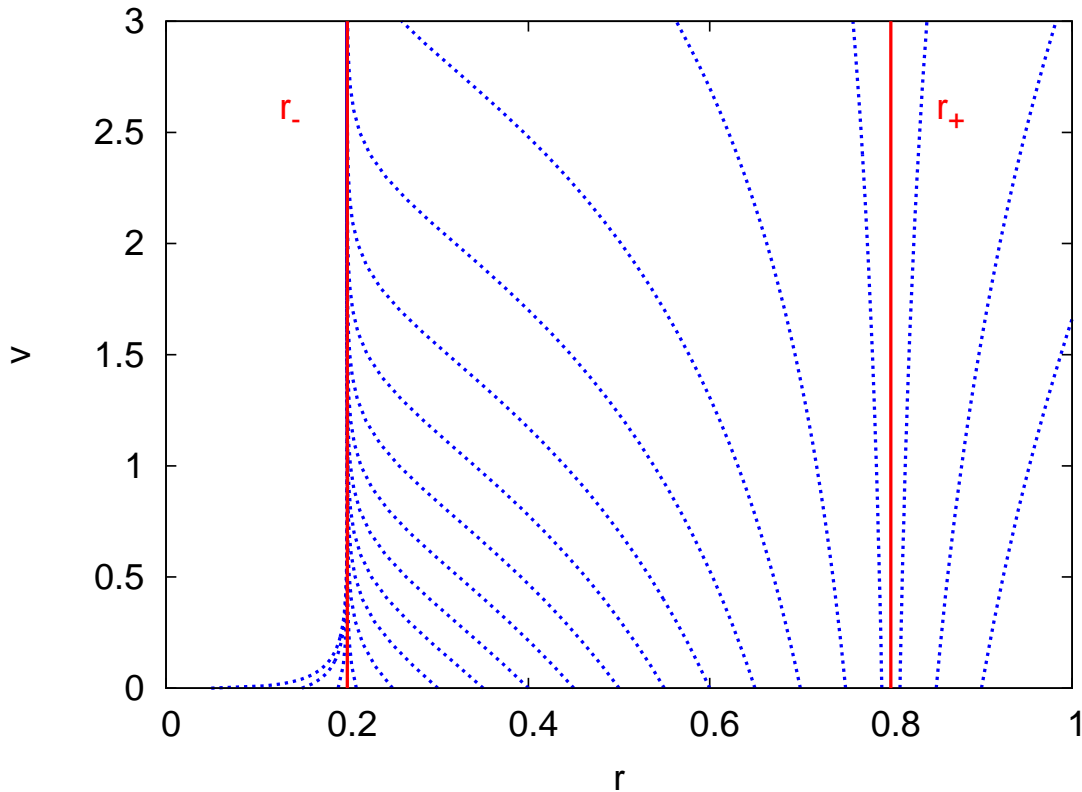}
\end{center}
\vspace{-0.5cm}
\caption{Example of $u$-constant null-geodesics (dotted lines) obtained from
  Eq.~(\ref{r2}) for the usual Reissner-Nordstr\"om case,
  corresponding to $B = -1/2$, $m = 0.5$, $q = 0.4$, $n = 4$,
  $\Lambda = 0$, and taking $r(u,v=0) = -u/2$. (See \cite{GS}). Here
  $r_+$ and $r_-$ (solid lines) are defined as usual as $r_{\pm} = m
  \pm \sqrt{m^2 - q^2}$. In the exterior region ($r>r_+$), the
  constant $u$ null geodesics reach ${\cal I}^+$, while in the
  interior region they are confined, giving origin to the typical
  black hole causal structure.} 
\label{fig1}
\end{figure}
The approach, which we will not reproduce here, allows us to qualitatively construct
conformal diagrams, identifying horizons and
singularities, and also to evaluate specific geometric quantities.
The main idea, however, is to solve Eq. (\ref{r2}) numerically as an
initial value in $v$, for constant $u$. In other words, we obtain
numerically $r(u,v)$ for $u$ constant, starting with a initial
condition 
\beq
\label{F(u)}
r(u,v_0) = F(u),
\eeq
where we must have $F'(u)\ne 0$, as can be seen from (\ref{B}). In analogy with the flat spacetime case, we choose here $F(u) = -\frac{u}{2}$.
Since the lines of constant $u$ (or $v$) are null
 geodesics for any metric in double-null coordinates, knowing $r(u,v)$
 for $u$ constant is enough, for instance, to construct the causal
 conformal diagrams. Figure (\ref{fig1}) depicts a simple example,
 corresponding to the usual Reissner-Nordstr\"om solution. 

\begin{figure}[!htb]
\vspace{-0.5cm}
\begin{center}
  \includegraphics[angle=270,width=1.0\linewidth]{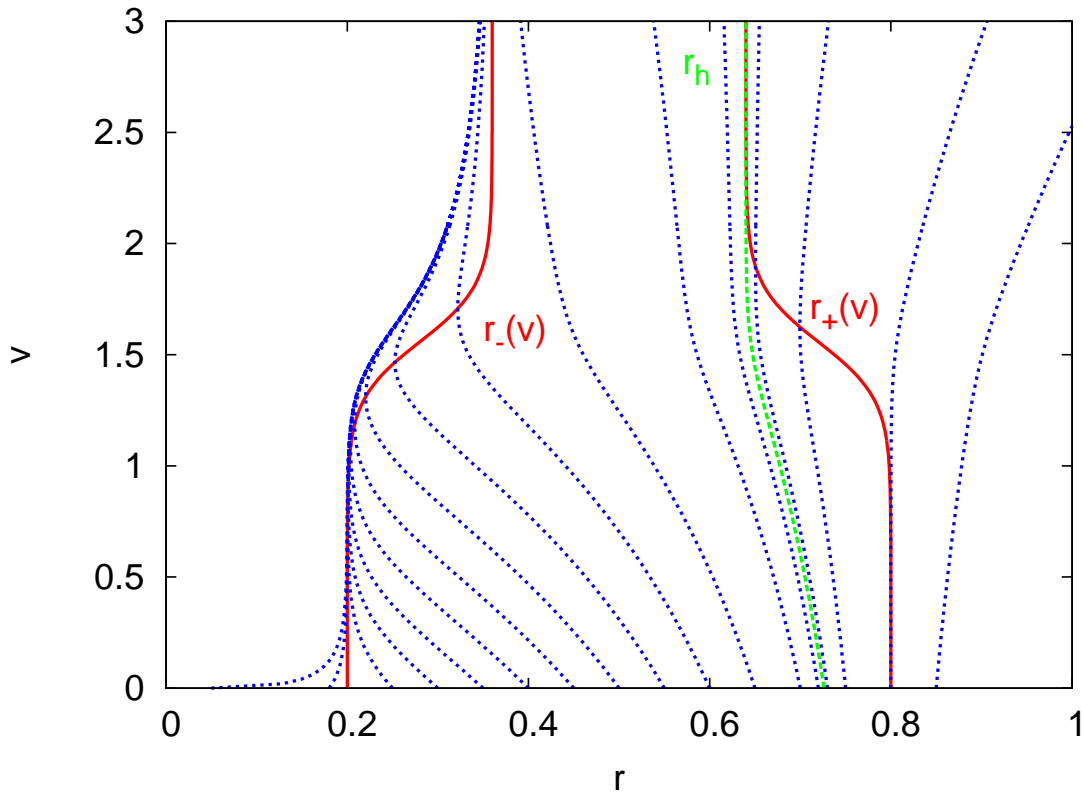}
\end{center}
\vspace{-0.5cm}
\caption{Example of $u$-constant null-geodesics (dotted lines) obtained from
  Eq.~(\ref{r2}), with the same parameter values used in Figure \ref{fig1} but now with a time dependent charge function given by $2q(v) = \left(q_f+q_i\right) + \left(q_f-q_i\right)\tanh \rho (v-v_m)$, with $q_i = 0.4$, $q_f = 0.48$, $\rho = 4.0$ and $v_m = 1.5$. We can see in this case that the $r_{\pm}$ (solid lines) are no longer constant, and the event horizon $r_h$ (dashed line) no longer coincides with $r_+$.} 
\label{fig2}
\end{figure}

\begin{figure}[!htb]
\vspace{-0.5cm}
\begin{center}
  \includegraphics[angle=270,width=1.0\linewidth]{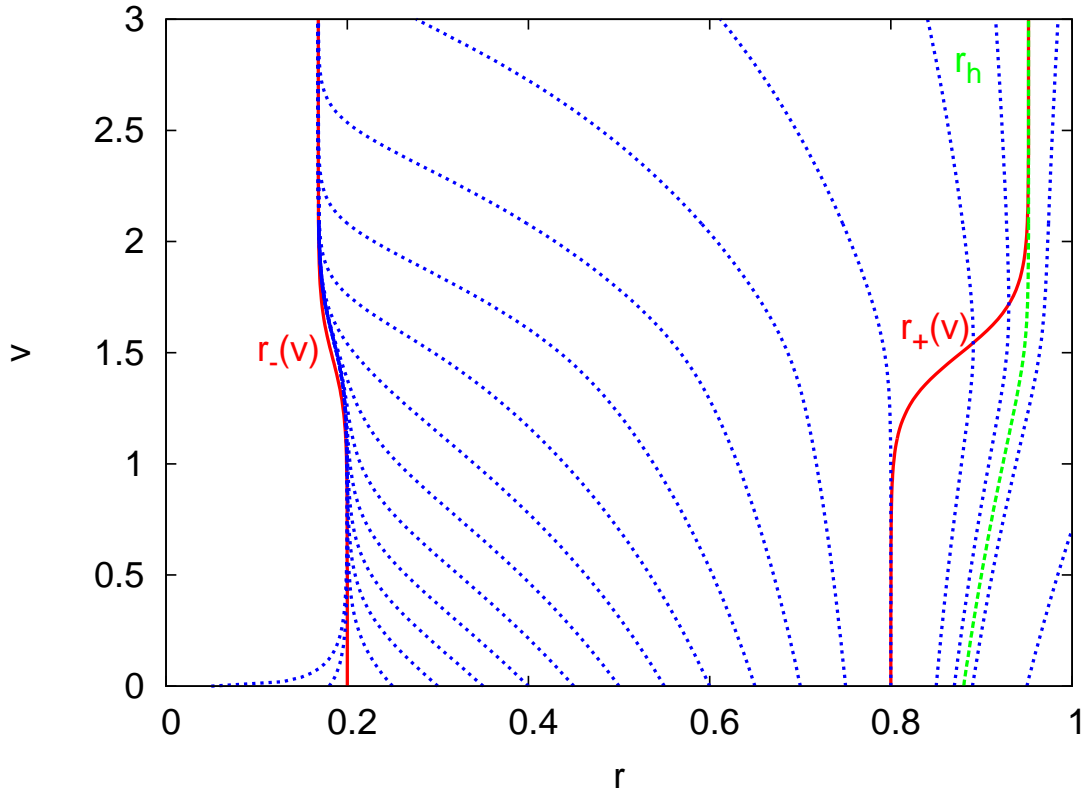}
\end{center}
\vspace{-0.5cm}
\caption{Same as Figure \ref{fig2}, but this time with $q = 0.4$ and $2m(v) = \left(m_f+m_i\right) + \left(m_f-m_i\right)\tanh \rho (v-v_m)$, with $m_i = 0.5$, $m_f = 0.56$, $\rho = 4.0$ and $v_m = 1.5$.} 
\label{fig3}
\end{figure}

\begin{figure}[!htb]
\vspace{-0.5cm}
\begin{center}
  \includegraphics[angle=270,width=1.0\linewidth]{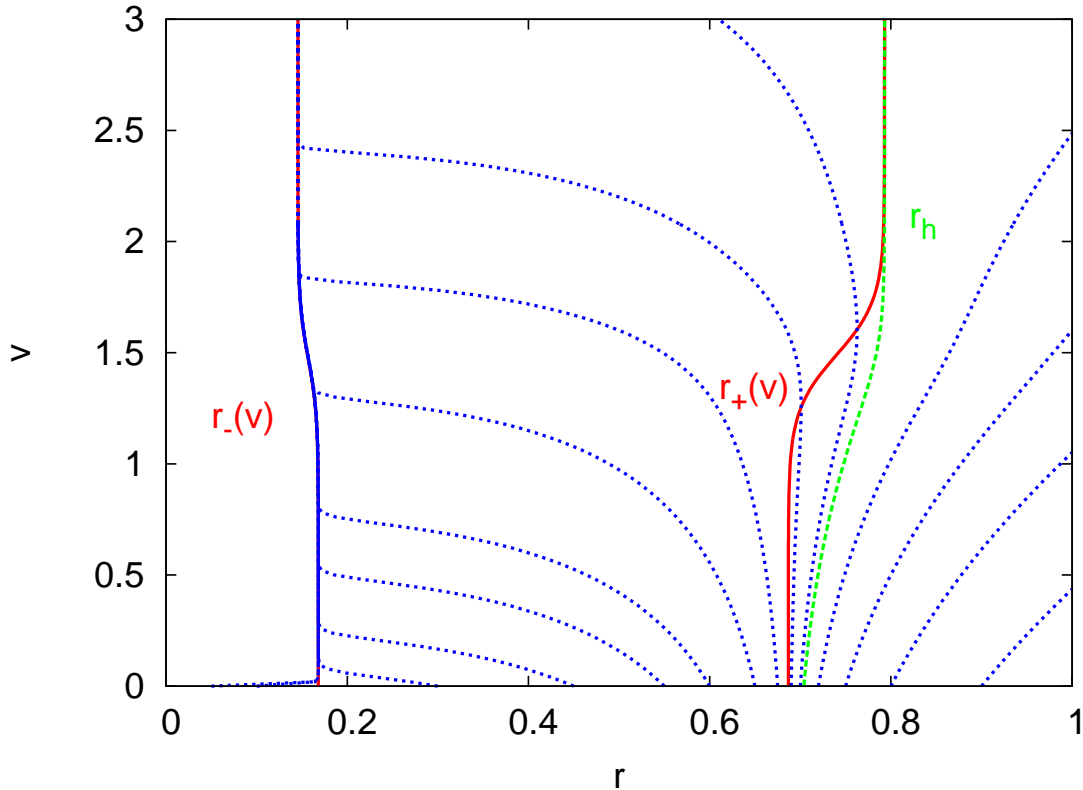}
\end{center}
\vspace{-0.5cm}
\caption{Same as Figure \ref{fig3}, but this time exploring a 5-dimensional solution ($n=5$), with $q = 0.2$, $m_i = 0.5$ and $m_f = 0.65$. The apparent horizons are now at $r\pm = \left[\frac{1}{2}\left(m \pm \sqrt{m^2 - \frac{4}{3} q^2}\right)\right]^{1/2}$ and therefore we must have $\sqrt{3}m > 2q$. The behavior is qualitatively the same as for the $n=4$ case.}
\label{fig4}
\end{figure}
\begin{figure}[!htb]
\vspace{-0.5cm}
\begin{center}
  \includegraphics[angle=270,width=1.0\linewidth]{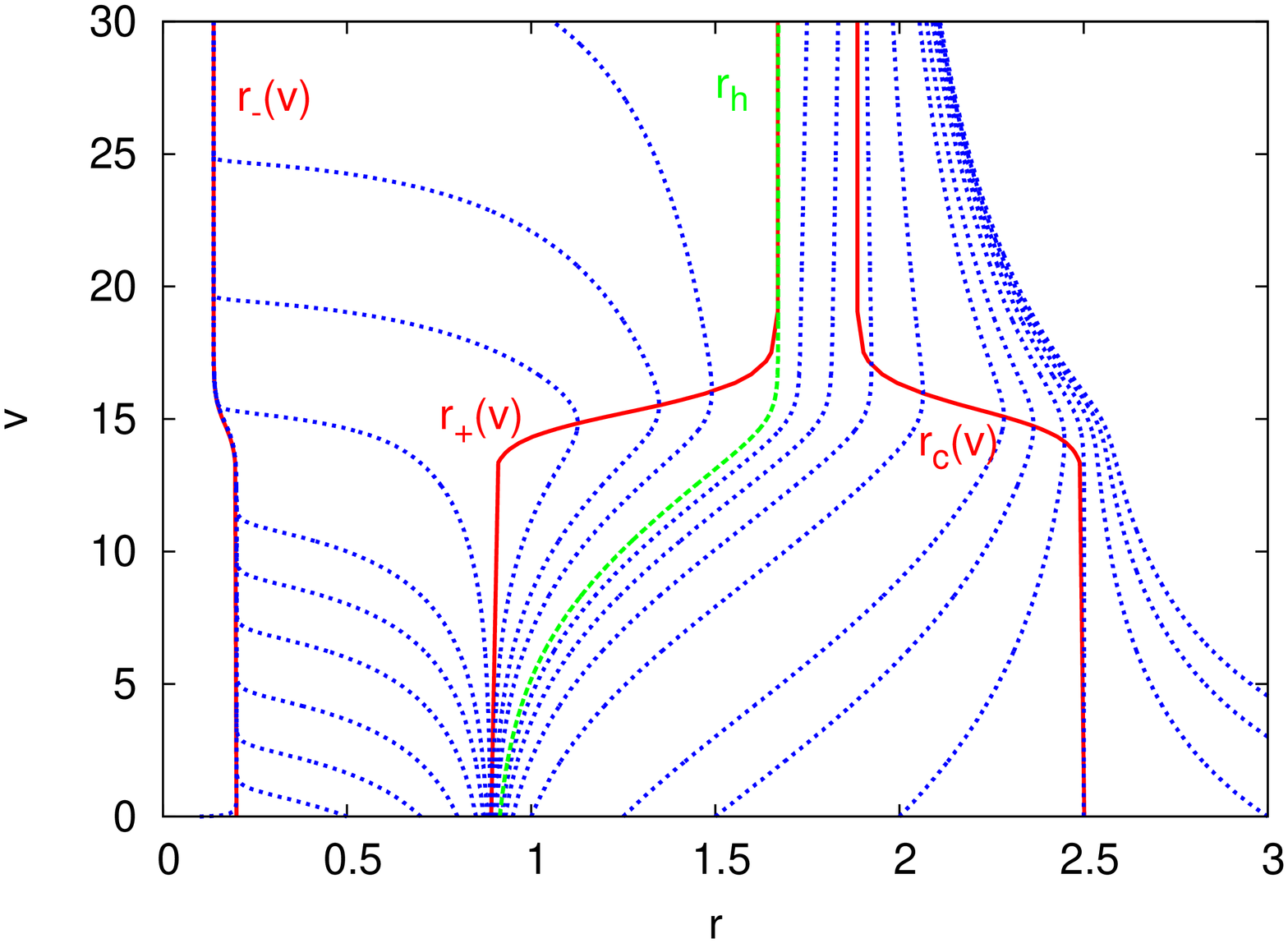}
\end{center}
\vspace{-0.5cm}
\caption{Same as Figure \ref{fig3}, but this time allowing a non-zero cosmological constant $\Lambda = 0.3$, with $q = 0.4$, $m_i = 0.5$, $m_f = 0.65$, $\rho = 1.0$ and $v_m = 15$. The striking feature in this case is the existence of three time-dependent apparent horizons $r_-, r_+$ and $r_c$, located at the positive roots of Eq. (\ref{r2}), which is a 4th order polynomial in this case.}
\label{fig5}
\end{figure}
Figures \ref{fig2} and \ref{fig3} present the behavior of the $u$-constant null-geodesics in two different time dependent cases: increasing charge function $q(v)$ and increasing mass function $m(v)$. Charge and mass variations produce opposite results for the horizons. Also, $r_-$ and $r_+$ are now time dependent, and the event horizon $r_h$ is no longer coincident with the apparent horizon $r_+$.

The case presented in Fig. 2 deserves a more detailed analysis. The charge increase is analogous to the mass evaporation studied in \cite{GS,Saa:2007ej}.
 
Following our discussion in Section \ref{sec2}, when $q_{,v} > 0$ we must have $B = +\frac{1}{2}$, in order to satisfy the weak energy condition (\ref{energycond}). Now there is a subtlety regarding the sign of $B$. We can see from Eq. (\ref{B}) that the sign of $f$ depends on the signs of $B$ and $r_{,u}$ (which is of negative sign with our choice of $F(u)$). Therefore, for $B = -\frac{1}{2}$, we have $f > 0$ and $\partial_u + \partial_v$ is timelike and $\partial_v - \partial_u$ is spacelike. However, for $B = +\frac{1}{2}$, $f$ has the opposite sign and the timelike and spacelike directions are now exchanged.

The transformation $(u,v) \to (v,-u)$ restores the temporal and spatial directions. Under this transformation, Eq. (\ref{r2}) (with $B = +\frac{1}{2}$) becomes

\begin{equation*}
\fl
-r_{,u} = -\frac{1}{2} \left(1 - \frac{2m(-u)}{(n-3)r^{n-3}} - \frac{2\Lambda}{(n-2)(n-1)}r^2 + \frac{2}{(n-2)(n-3)}\frac{q^2(-u)}{r^{2(n-3)}}
\right),
\end{equation*}

which is formally identical to a case with $B = -\frac{1}{2}$, mass function $m(-u)$ and charge function $q(-u)$. 
Therefore, the
transformed equation results in a situation with {\it decreasing charge}, that is, a time reversal of the original situation with increasing charge.
Stated in a different way, this means that the case with increasing charge and $B = +\frac{1}{2}$ must be interpreted as the time reversal of the case with decreasing charge and $B = -\frac{1}{2}$ (with both cases satisfying the weak energy condition).

However, in order to describe an actual charge increase (equivalent to an evaporation), we need to violate the weak energy condition, since there are no classical processes that can lead to black hole evaporation. In order to do that, we deliberately choose $B = -\frac{1}{2}$ together with $q_{,v} > 0$. The weak energy condition is violated, and the resulting evaporation process is shown in Figure 2.

In Figure \ref{fig4} we show an example of a higher dimensional spacetime with $n=5$. The results are qualitatively the same as for the 4-dimensional cases we presented in Figs. \ref{fig2} and \ref{fig3}. In Figure \ref{fig5} we have a qualitatively different behavior, due to the appearance of a cosmological horizon ($\Lambda \ne 0$).

The calculation of the apparent horizons $r\pm(v)$ shown in Figures \ref{fig1}-\ref{fig5} follows the semi-analytic method described in refs. \cite{GS,Saa:2007ej}. We can see that the curves defined by the vanishing of the r.h.s of Eq. (\ref{r2}) always describe the frontier between two regions of the $(v,r)$ plane where the solutions of Eq. (\ref{r2}) have distinct behaviors. For $r_{,v} < 0$, the null geodesics approach the singularity, whereas for $r_{,v} > 0$, the null geodesics tend to escape from the singularity.

The determination of the event horizon $r_h$ is done numerically by inspection of the initial value $r(u,v_0)$. The event horizon is found as the last geodesic that escapes towards infinity and does not fall into the singularity. Note that in the case of increasing charge (or, equivalently, decreasing mass) there are null geodesics that escape towards infinity even though they were initially inside the apparent horizon $r_+$.

\section{Final discussion}
\label{sec4}

We have presented a formulation in double-null coordinates of the most general Vaidya metric: $n$-dimensional, with varying mass and/or charge and cosmological constant $\Lambda$. 

By exploring the numerical solutions of Eq.~(\ref{r2}), we were able to highlight some interesting features of the behavior of time-dependent horizons in multiple-horizon spacetimes. The $u$-constant geodesics can be used to track the time dependent event and Cauchy horizons, that no longer coincide with $r_+$ and $r_-$.

The formulation presented here for the Einstein Eqs.~(\ref{B})-(\ref{r2}) was recently used in a quasinormal mode analysis of the Vaidya metric \cite{Cecilia}, and provided the framework needed to obtain the quasinormal frequencies with sufficient accuracy to verify their nonstationary behavior.

\ack

This work was supported by CNPq, FAPESP and the Max Planck Society. 

\appendix
\section*{Appendix}
\setcounter{section}{1}

Here we present a short discussion and the results for the Einstein equations (\ref{eq1})-(\ref{eq4}) for the $n=3$ case, generalizing the discussion of \cite{Saa:2007ej}. For $n=3$, Eq.~(\ref{eq1}) is still valid and can be integrated to give

\beq
f = 2Br_{,u}\,,
\label{ap_f}
\eeq
same as Eq.~(\ref{B}). 

Taking $n=3$, Eq.~(\ref{eq4}) reads
\beq
rr_{,uv} - (\Lambda r^2 + q^2)f = 0\,,
\label{ap1}
\eeq
and we can use Eq.~(\ref{ap_f}) to integrate Eq.~(\ref{ap1}) and obtain
\beq
r_{,v} = -B(-m - \Lambda r^2 - 2q^2\ln r)\,,
\label{ap_r2}
\eeq
which is the $n=3$ version of Eq.~(\ref{r2}). Here $m(v)$ is an integration function that has the same interpretation as before, that is, the mass of the solution. Compare (\ref{ap_r2}) with the charged BTZ black hole \cite{BTZ}.

From Eq.~(\ref{eq2}) with $n=3$, we have
\beq
h = \frac{r_{,v}}{r}\frac{f_{,v}}{f} - \frac{r_{,vv}}{r}\,,
\eeq
and using Eqs.~(\ref{ap_r2}) and (\ref{ap_f}) to get
\beq
f_{,v} = 2B\left(\Lambda r + \frac{q^2}{r}\right)f\,,
\eeq
we obtain the $n=3$ version of Eq.~(\ref{eq2}):
\beq
h = -B\left( \frac{m_{,v}}{r} + \frac{4\ln r}{r}qq_{,v}\right)\,.
\label{ap_h}
\eeq

We also note here that the $n=3$ version of Eq.~(\ref{eq3}),
\beq
 \frac{f_{,u}f_{,v}}{f^2} - \frac{f_{,uv}}{f} - \frac{r_{,uv}}{r} =  -2\Lambda f\,, 
\eeq
together with Eq.~(\ref{ap_f}) still reproduce Eq.~(\ref{A}) with $n=3$.

So finally we can conclude that the Einstein equations (\ref{eq1})-(\ref{eq4}), which in the $n \ge 4$ case are equivalent to Eqs.~(\ref{B})-(\ref{r2}), are equivalent in the $n=3$ case to Eqs.~(\ref{ap_f}), (\ref{ap_r2}) and (\ref{ap_h}).

\end{document}